\documentclass[runningheads]{llncs}
\usepackage{graphicx}
\usepackage{color}
\usepackage{booktabs}
\usepackage{multirow}
\usepackage{amsmath}
\usepackage{amssymb}
\usepackage{hyperref}
\usepackage{makecell}
\usepackage{chngpage}
\usepackage[misc]{ifsym}
\usepackage{mathtools}
\usepackage{wrapfig}
\usepackage{makecell}

\newcommand{\ours}{\texttt{Com-BrainTF}{}}

\newcommand{\eg}[0]{{\textit{e.g.},}}

\begin{document}
\title{Community-Aware Transformer for Autism Prediction in fMRI Connectome}
%
%

\author{Anushree Bannadabhavi\inst{1}
\and
Soojin Lee\inst{2} 
\and
Wenlong Deng
\inst{1}
\and Xiaoxiao Li\inst{1}
}
\authorrunning{A. Bannadabhavi et al.}

%
\institute{The University of British Columbia, Vancouver, BC, Canada
\email{anushree.bannadabhavi@gmail.com}\\}

\institute{
Department of Electrical and Computer Engineering,\\The University of British Columbia, Vancouver, BC V6Z 1Z4, Canada
\and 
Department of Medicine, \\ The University of British Columbia Vancouver, BC V6Z 1Z4, Canada}


\maketitle              
\begin{abstract}
Autism spectrum disorder(ASD) is a lifelong neurodevelopmental condition that affects social communication and behavior. Investigating functional magnetic resonance imaging (fMRI)-based  brain functional connectome can aid in the understanding and diagnosis of ASD, leading to more effective treatments. The brain is modeled as a network of brain Regions of Interest (ROIs), and ROIs form communities and knowledge of these communities is crucial for ASD diagnosis. On the one hand, Transformer-based models have proven to be highly effective across several tasks, including fMRI connectome analysis to learn useful representations of ROIs. On the other hand, existing transformer-based models treat all ROIs equally and overlook the impact of community-specific associations when learning node embeddings. To fill this gap, we propose a novel method, \ours{}, a hierarchical local-global transformer architecture that learns intra and inter-community aware node embeddings for ASD prediction task. Furthermore, we avoid over-parameterization by sharing the local transformer parameters for different communities but optimize unique learnable prompt tokens for each community. Our model outperforms state-of-the-art (SOTA) architecture on ABIDE dataset and has high interpretability, evident from the attention module. Our code is available at \href{https://github.com/ubc-tea/Com-BrainTF}{https://github.com/ubc-tea/Com-BrainTF}. \let\thefootnote\relax\footnote{This work is supported in part by the Natural Sciences and Engineering Research Council of Canada (NSERC) and Compute Canada.\\ Corresponding author: Xiaoxiao Li (\email{xiaoxiao.li@ece.ubc.ca})} 

\end{abstract}
\section{Introduction}
Autism spectrum disorder (ASD) is a developmental disorder that affects communication, social interaction, and behaviour \cite{Wing1979}. ASD diagnosis is challenging as there are no definitive medical tests such as a blood tests, and clinicians rely on behavioural and developmental assessments to accurately diagnose ASD. As a more objective alternative measurement, neuroimaging tools, such as fMRI has been widely used to derive and validate biomarkers associated with ASD \cite{Goldani2014}. Conventionally, fMRI data is modelled as a functional connectivity (FC) matrix, \eg{} by calculating Pearson correlation coefficients of pairwise brain ROIs to depict neural connections in the brain. FC-based brain connectome analysis is a powerful tool to study connectivity between ROIs and their impact on cognitive processes, facilitating diagnosis and treatment of neurological disorders including ASD \cite{Kaiser2010,Price2014}.

Deep learning (DL) models have led to significant advances in various fields including fMRI-based brain connectome analysis. For instance, BrainNetCNN\cite{KAWAHARA20171038} proposes a unique convolutional neural network (CNN) architecture with special edge-to-edge, edge-to-node, and node-to-graph convolutional filters. Considering brain network's non-euclidean nature, graph neural networks (GNNs) \cite{veličković2018graph} have emerged as a promising method for analyzing and modelling brain connectome by constructing graph representation from FC matrices~\cite{veličković2018graph,LI2021102233,Fbnetgen,cui2022interpretable}. More recently, transformers~\cite{NIPS2017_3f5ee243} have been utilized for brain connectome analysis as they can learn long-range interaction between ROIs without a predefined graph structure. Brain Network Transformer (BNT)~\cite{kan2022bnt} has transformer encoders that learn node embeddings using a pearson correlation matrix and a readout layer that learns brain clusters. BNT~\cite{kan2022bnt} shows that transformer-based methods outperform CNN and GNN models for fMRI-based classification.

It is worth noting that our brain is comprised of functional communities \cite{vandenHeuvel2010} (often referred to as functional networks), which are groups of ROIs that perform similar functions as an integrated network \cite{Smith2009,Smith2013}. These communities  are highly reproducible across different studies \cite{Chen2008,Damoiseaux2006} and are essential in understanding the functional organization of the brain \cite{vandenHeuvel2009}. Their importance extends beyond basic neuroscience, as functional communities have been shown to be relevant in understanding cognitive behaviours \cite{vandenHeuvel2010}, mental states \cite{Geerligs2015}, and neurological and psychiatric disorders \cite{Rosazza2011,Canario2021}. For example, numerous studies have found that individuals with ASD exhibit abnormalities in default mode network (DMN) \cite{Padmanabhan2017}, which is characterized by a combination of hypo- and hyper-connectivity with other functional networks. ASD is also associated with lifelong impairments in attention across multiple domains, and significant alterations in the dorsal attention network (DAN) and DMN have been linked to these deficits \cite{Farrant2015}. Such prior knowledge of functional relationships between functional communities, can be highly beneficial in predicting ASD. However, existing DL models often overlook this information, resulting in sub-optimal brain network analysis.

To address this limitation, we propose a novel community-aware transformer for brain network analysis, dubbed \ours{}, which integrates functional community information into the transformer encoder. \ours{} consists of a hierarchical transformer with a local transformer that learns community-specific embeddings, and a global transformer that fuses the whole brain information. The local transformer takes FC matrices as input with its parameters shared across all communities, while, personalized prompt tokens are learnt to differentiate the local transformer embedding functions. The local transformer's output class tokens and node embeddings are passed to the global transformer, and a pooling layer summarizes the final prediction. Our approach enhances the accuracy of fMRI brain connectome analysis and improves understanding of the brain's functional organization. Our key contributions are:\\ 
\noindent 1. We propose a novel local-global hierarchical transformer architecture to efficiently learn and integrate community-aware ROI embeddings for brain connectome analysis by utilizing both ROI-level and community-level information. \\
\noindent 2.  We avoid over-parameterization by sharing the local transformer parameters and design personalized learnable prompt tokens for each community. \\
\noindent 3.  We prove the efficacy of \ours{} with quantitative and qualitative experiment results. Our visualization demonstrates the ability of our model to capture functional community patterns that are crucial for ASD vs. Healthy Control (HC) classification.

\section{Method}

\subsection{Overview}

\begin{figure}[t]
\includegraphics[width=\textwidth]{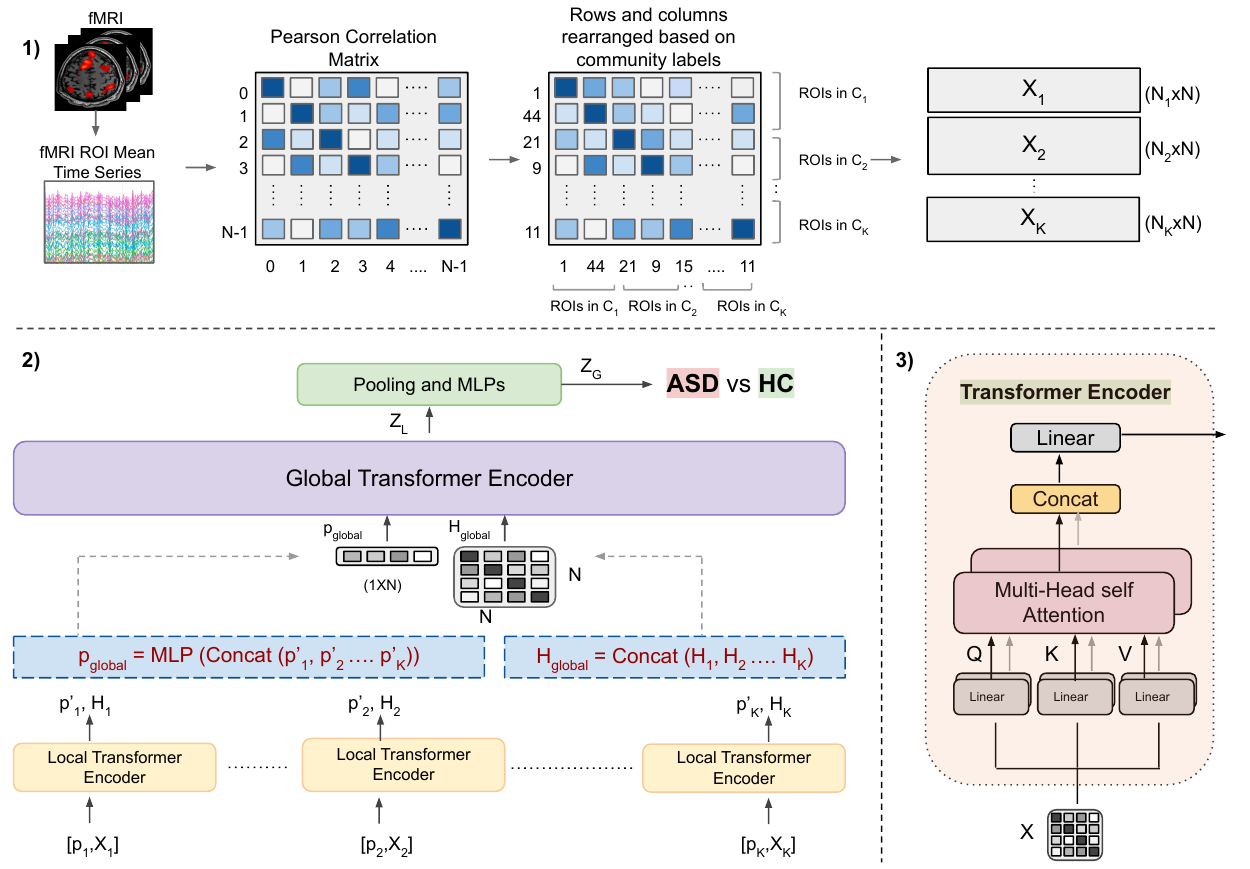}
\caption{1) fMRI images parcellated by an atlas to obtain the Functional connectivity matrix for 'N' ROIs. Rows and columns are rearranged based on community labels of each ROI to obtain matrices $X_1, X_2 ... X_K$, inputs to local transformer 2) Overview of our local-global transformer architecture 3) Transformer encoder module} \label{fig:architecture}
\vspace{-5mm}
\end{figure}

\noindent\textbf{Problem Definition}
In brain connectome analysis, we first parcellate the brain into $N$ ROIs based on a given atlas. FC matrix is constructed by calculating the Pearson correlation coefficients between pairs of brain ROIs based on the strength of their fMRI activations. Formally, given a brain graph with $N$ number of nodes, we have a symmetric FC matrix $X \in \mathbb{R}^{N \times N}$. Node feature vector of ROI $j$ is defined as the $j^{th}$ row or column of this matrix. Given $K$ functional communities and the membership of ROIs, we 
rearrange the rows and columns of the FC matrix, resulting in $K$ input matrices $\{X_1, X_2, \dots, X_K\}$.  $X_k \in \mathbb{R}^{N_k \times N}$ is viewed as a $N_k$ length sequence with $N$ dimensional tokens and $\sum_k N_k = N$ (Fig.~\ref{fig:architecture}(1)).
This process helps in grouping together regions with similar functional connectivity patterns, facilitating the analysis of inter-community and intra-community connections. \ours{} inputs $X_k$ to the community $k$-specific local transformer and outputs $N$ dimensional tokens $H_k \in \mathbb{R}^{N_k \times N}$. 
A global transformer learns embedding  $H = [H_1, \dots, H_k] \mapsto Z_L \in \mathbb{R}^{N \times N}$, followed by a pooling layer and multi-layer perceptrons (MLPs) to predict the output.\\

\noindent\textbf{Overview of our Pipeline}
Human brain connectome is a hierarchical structure with ROIs in the same community having greater similarities compared to inter-community similarities. Therefore, we designed a local-global transformer architecture(Fig.~\ref{fig:architecture}(2)) that mimics this hierarchy and efficiently leverages community labels to learn community-specific node embeddings. This approach allows the model to effectively capture both local and global information. Our model has three main components: a local transformer, a global transformer and a pooling layer details of which are discussed in the following subsections.

\subsection{Local-global transformer encoder}
\label{sec:loc-glo}

The transformer encoder\cite{NIPS2017_3f5ee243} (Fig.~\ref{fig:architecture}(3)) is a crucial component of both local and global transformers. The transformer encoder takes as input an FC matrix and has a multi-head-attention module to capture interdependencies between nodes using attention mechanism. The learnt node feature matrix $H_i$ is given by
\begin{equation}
H_{i} = (\parallel_{m=1}^M h^m) W_O,
\end{equation}
\begin{equation}
h^m = \mbox{softmax}\Bigg(\frac{Q^m(K^m)^T}{\sqrt{d_k^m}}\Bigg)V^m, \mbox{where,}
\end{equation}
$Q^m = W_QX_i', K^m = W_KX_i'$, $V^m = W_VX_i'$, $X_i' = [p_i, X_i]$, $M = $ number of attention heads, $\parallel$ is concatenation operator, $W_Q, W_K, W_V$, $W_O$ are model parameters.\\

\noindent\textbf{Local Transformer:}
For optimal analysis of fMRI brain connectomes, it is important to incorporate both functional community information and node features. Therefore, we introduce the knowledge of community labels to the network by grouping node features based on community labels producing inputs $\{X_1, X_2, \dots, X_K\}$. However, separate local transformers for each input would result in a significant increase in model parameters. Hence, we use the same local transformer for all inputs, but introduce unique learnable personalized prompt tokens $\{p_1, p_2, \dots, p_K\}$, where $p_i \in \mathbb{R}^{1 \times N}$, that learn to distinguish between node feature matrices of each community and therefore avoid over-parameterization of the model. Previous transformer-based models like BNT\cite{kan2022bnt} do not have any prompt tokens and only use the FC matrix as input. Using attention mechanism, pairwise connection strength between ROIs is learnt, producing community-specific node embeddings $H_i$ and prompt tokens $p^{'}_i$,

\begin{equation}
p^{'}_i, H_i = \mbox{LocalTransformer}([p_i, X_i]) \mbox{ where, } i \in [1,2...K].
\end{equation}

\noindent\textbf{Global Transformer:} On obtaining community-specific node embeddings and prompt tokens from the local transformer, it is essential to combine this information and design a module to learn the brain network at a global level. Therefore, we introduce a global transformer encoder to learn inter-community dependencies. Input to the global transformer is the concatenated, learnt node feature matrices from the local transformer and a prompt token (Fig.~\ref{fig:architecture}(2)). Prompt tokens learnt by the local transformer contain valuable information to distinguish different communities and thus are concatenated and passed through an MLP to obtain a prompt token input for the global transformer as follows:
\begin{equation}
p_{global} = \mbox{MLP }(\mbox{Concat }(p^{'}_1, p^{'}_2 \dots  p^{'}_{K})),
\end{equation}
\begin{equation}
H_{global} = \mbox{Concat}(H_1, H_2, \dots , H_{K}),
\end{equation}
\begin{equation}
p^{'}, Z_L = \mbox{GlobalTransformer}([p_{global}, H_{global}]).
\end{equation}
The resulting attention-enhanced, node embedding matrix $Z_L$ is then passed to a pooling layer for further coarsening of the graph. Extensive ablation studies are presented in section 3.3 to justify the choice of inputs and output.

\subsection{Graph Readout Layer}

The final step involves aggregating global-level node embeddings to obtain a high-level representation of the brain graph. We use OCRead layer \cite{kan2022bnt} for aggregating the learnt node embeddings. OCRead initializes orthonormal cluster centers $E \in \mathbb{R}^{K \times N}$ and softly assigns nodes to these centers. 
Graph level embedding $Z_G$ is then obtained by $Z_G = A^\top Z^L$, where $A \in \mathbb{R}^{K \times N}$is a learnable assignment matrix computed by OCRead. $Z_G$ is then flattened and passed to an MLP for a graph-level prediction. The whole process is supervised with Cross-Entropy (CE) loss.

\section{Experiments}

\subsection{Datasets and Experimental Settings}

\noindent\textbf{ABIDE} is a collection of resting-state functional MRI (rs-fMRI) data from 17 international sites \cite{ABIDE} with Configurable Pipeline for the Analysis of Connectomes (CPAC), band-pass filtering (0.01 - 0.1 Hz), no global signal regression, parcellated by Craddock 200 atlas\cite{Craddock2012}. ROIs of ABIDE dataset belong to either of the eight functional communities namely, cerebellum and subcortical structures (CS \& SB), visual network (V), somatomotor network (SMN), DAN, ventral attention network (VAN), limbic network (L), frontoparietal network (FPN) and DMN, following the assignments in \cite{yeo2011kriene}. ABIDE has fMRI data of 1009 subjects, 51.14\% of whom were diagnosed with ASD. Pearson correlation matrix is computed using mean time series of ROIs. Stratified sampling strategy\cite{kan2022bnt} is used for train-validation-test data split.

\noindent\textbf{Experimental Settings:} All models are implemented in PyTorch and trained on NVIDIA V100 with 8GB memory. For both local and global transformers, the number of attention heads is equal to the number of communities. We used Adam optimizer with an initial learning rate of $10^{-4}$ and a weight decay of $10^{-4}$. The train/validation/test data spilt ratio is 70:10:20, and the batch size is 64.
Prediction of ASD vs. HC is the binary classification task. We use AUROC, accuracy, sensitivity, and specificity on the test set to evaluate performance. Models are trained for 50 epochs and we use an early stopping strategy. The epoch with the highest AUROC performance on the validation set is used for performance comparison on the test set.
\subsection{Quantitative and Qualitative Results}
\label{sec:quant}
\noindent\textbf{Comparison with Baselines (Quantitative results)}
Following the comparison method mentioned in BNT\cite{kan2022bnt}, we compare the performance of our model with three types of baselines (i) Comparison with transformer-based models - BNT (ii) Comparison with neural network model on fixed brain network - BrainNetCNN\cite{KAWAHARA20171038} (iii) Comparison with neural network models on learnable brain network - FBNETGEN\cite{Fbnetgen}. Note that the selected baselines are reported to be the best among other alternatives in the same category on ABIDE dataset by BNT\cite{kan2022bnt}. As seen in Table.\ref{tab:perf}, our model outperforms all the other architectures on three evaluation metrics. In contrast to other models, \ours{} is hierarchical, similar to the brain's functional organization and therefore is capable of learning relationships within and between communities. 

\begin{table}[t]
\caption{Performance comparison with baselines (Mean $\pm$ standard deviation)}\label{tab:perf}
\centering
\begin{tabular}{p{0.22\textwidth} p{0.18\textwidth} p{0.18\textwidth} p{0.18\textwidth} p{0.14\textwidth}}
\hline
\textbf{Model} &  \textbf{AUROC} & \textbf{Accuracy} & \textbf{Sensitivity} & \textbf{Specificity}\\
\hline
BrainNetTF\cite{kan2022bnt} &  78.3 $\pm$ 4.4 &  68.1 $\pm$ 3.1 & 78.1 $\pm$ 10.0 & 58.9 $\pm$ 12.0\\
BrainNetCNN\cite{KAWAHARA20171038} & 74.1 $\pm$ 5.1 &  67.5 $\pm$ 3.1 & 65.3 $\pm$ 4.3 & \textbf{69.6} $\pm$ \textbf{4.1}\\
FBNETGNN\cite{Fbnetgen} &  72.5 $\pm$ 8.3 &  64.9 $\pm$ 8.9 & 60.9 $\pm$ 11.3 & 67.5 $\pm$ 13.1\\
\textbf{\ours{}} & \textbf{79.6 $\pm$ 3.8} &  \textbf{72.5 $\pm$ 4.4} & \textbf{80.1 $\pm$ 5.8} & 65.7 $\pm$ 6.4\\
\hline
\end{tabular}
\end{table}


\begin{figure}[t]
\includegraphics[width=\textwidth]
{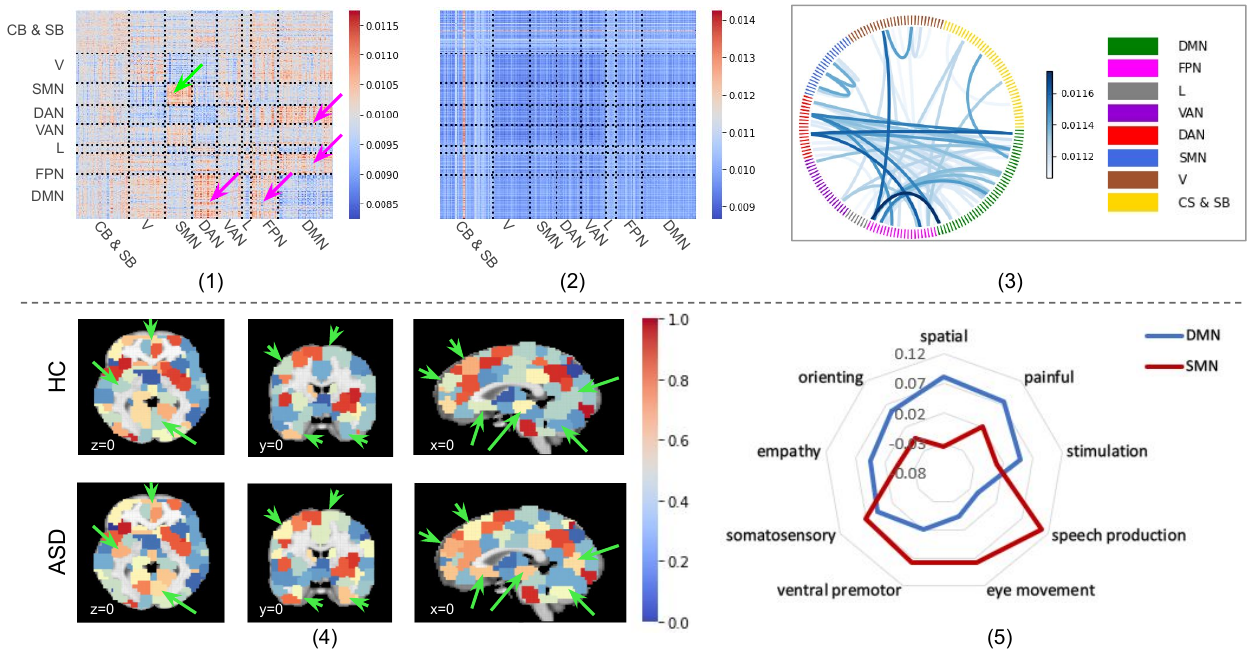}
\caption{(1) Attention matrix of the global transformer of Com-BrainTF highlighting prominent communities that influence ASD prediction (2) Attention matrix of the first transformer encoder module of BNT (3) Chord plot of the top 1\% of attention values (4) Neurosynth results generated from averaged prompt vectors (5) Correlation between functional communities and functional keywords decoded by Neurosynth~\cite{Yarkoni2011}.} \label{fig:interpretation}
\vspace{-5.5mm}
\end{figure}

\noindent\textbf{Interpretibility of Com-BrainTF (Qualitative results)}
We perform extensive experiments to analyze the interpretability of \ours{}. The following results are discussed in detail:\\
\noindent\textbf{1. Interpretation of the learned attention matrix}: Fig.~\ref{fig:interpretation}(1) and Fig.~\ref{fig:interpretation}(2) show the averaged attention matrices of \ours{} and BNT\cite{kan2022bnt} respectively. These attention matrices are obtained by averaging attention scores over correctly classified test data.  In comparison to BNT, our learned attention scores clearly highlight the communities important for ASD prediction. Specifically, attention scores in the SMN region (green arrow) are high indicating that within SMN, connectivity plays an essential role in ASD prediction. This aligns with prior studies on autism which have shown reduced FC within SMN \cite{Ilioska2022}, reflecting altered sensory and motor processing \cite{MARCO2011,Gowen2012,Mostofsky2011}. Additionally, attention scores in DMN, DAN, and FPN regions were found to be high (magenta arrows) suggesting that the functional connectivity between DMN and DAN and between DMN and FPN are also crucial for ASD prediction which is in line with previous studies that report abnormalities in DMN\cite{Padmanabhan2017} and its functional connectivity with other functional networks including DAN and FPN.  These connections can be better visualized from the chord plot in Fig.~\ref{fig:interpretation}(3). To summarize, \ours{} is able to correctly identify functional communities associated with ASD and the results are consistent with neuroscience findings.\\
\noindent\textbf{2. Ability of the prompt's attention vector to differentiate between ASD and HC subjects}: We investigate the first row of the learned attention matrix, namely the attention of ROIs corresponding to the prompt. The ROI-wise normalized attention scores shown in Fig.~\ref{fig:interpretation}(4) are generated using averaged prompt vectors over correctly classified test data, where red to blue indicate attention scores from high to low (dark red:1 to dark blue: 0). As evident from the figure, clear differences can be seen in the important brain regions among ASD and HC subjects. Additionally, most differences in the prompt vector embeddings are seen in DMN and SMN (supporting figures in the appendix), consistent with previous neuroscience literature.\\
\noindent\textbf{3. Meta-analysis of the important functional networks that influence ASD prediction}: DMN and SMN networks have been found to be crucial for ASD prediction based on Fig.~\ref{fig:interpretation}(4). Using the difference of the prompt vector values in those regions between ASD and HC subjects, Fig.~\ref{fig:interpretation}(5) was generated based on Meta-analysis using Neurosynth\footnote{The meta-analytic framework (\href{www.neurosynth.org}{www.neurosynth.org}) provides the posterior probability $P$(Feature$\mid$Coordinate) for psychological features (i.e., word or phrase) at a given spatial location based on neuroscience literature.} database\cite{Yarkoni2011}. Regions in DMN were found to be associated with trait keywords such as empathy, painful, stimulation, orienting and spatial, whereas regions in the SMN showed higher correlations with eye movement, ventral premotor, somatosensory and speech production.
\subsection{Ablation studies}
We conduct ablation studies to justify the designs of input and output pairing of the global transformer that facilitates effective learning of node embeddings, consequently resulting in superior performance.

\noindent\textbf{Input: node features vs. class tokens of local transformers.} We evaluated two input possibilities for the global transformer: (i) use only prompt tokens from the local transformer (ii) incorporate both prompt tokens and updated node features. Table.\ref{tab:ablation} reveals that prompt tokens alone are unable to capture the complete intra-functional connectivity and negatively affect performance. 

\noindent\textbf{Output: Cross Entropy loss on the learned node features vs. prompt token.} In comparison to vision models like ViT\cite{VIT}, which only use the class token (similar to the prompt token in our structure) for classification, we use output node embeddings from the global transformer for further processing steps. We justify this design by conducting a performance comparison experiment: (i) use learnt node features as the global transformer output (ii) use only the prompt token as output. The results (Table.\ref{tab:ablation}) demonstrated a significant decrease in performance when prompt token alone was used. This is expected since brain networks are non-euclidean structures and the learnt node features capture more information about the underlying graph, making them essential for accurately capturing the inter-community dependencies.
\begin{table}[t]
\caption{Ablation study: Examining different input and output possibilities}\label{tab:ablation}
\centering
\begin{tabular}{p{0.32\textwidth} p{0.14\textwidth} p{0.14\textwidth} p{0.14\textwidth} p{0.14\textwidth}}
\hline
\multicolumn{5}{c}{\textbf{Different inputs to the global transformer}}\\
\hline
Inputs &  AUROC & Accuracy & Sensitivity & Specificity\\
\hline
Prompt tokens only &  73.2 $\pm$ 4.5 &  65.7 $\pm$ 5.8 & \textbf{88.8} $\pm$ \textbf{4.1} & 43.7 $\pm$ 9.1\\
Node features and prompt token & \textbf{79.6} $\pm$ \textbf{3.8} &  \textbf{72.5} $\pm$ \textbf{4.4} & 80.1 $\pm$ 5.8 & \textbf{65.7} $\pm$ \textbf{6.4}\\
\hline

\hline
\multicolumn{5}{c}{\textbf{CE loss on node features vs prompt token}}\\
\hline
Outputs &  AUROC & Accuracy & Sensitivity & Specificity\\
\hline
Prompt token &  71.9 $\pm$ 4.6 &  66.1 $\pm$ 3.6 & \textbf{80.4} $\pm$ \textbf{8.2} & 51.5 $\pm$ 7.6\\
Node features & \textbf{79.6} $\pm$ \textbf{3.8} &  \textbf{72.5} $\pm$ \textbf{4.4} & 80.1 $\pm$ 5.8 & \textbf{65.7} $\pm$ \textbf{6.4}\\
\hline
\end{tabular}
\vspace{-3mm}
\end{table}
\section{Conclusion}
In this work, we introduce \ours{}, a hierarchical, community-aware local-global transformer architecture for brain network analysis. Our model learns intra- and inter-community aware node embeddings for ASD prediction tasks. With built-in interpretability, \ours{} not only outperforms SOTA on the ABIDE dataset but also detects salient functional networks associated with ASD. We believe that this is the first work leveraging functional community information for brain network analysis using transformer architecture. Our framework is generalizable for the analysis of other neuroimaging modalities, ultimately benefiting neuroimaging research. Our future work includes investigating alternate variants for choosing different atlases and community network parcellations.

%
%
%
\bibliographystyle{splncs04}
\bibliography{reference}

\newpage
\appendix

\section{Supplementary Materials}

\subsection{Variations on the Number of Prompts}
\begin{table}[]
\caption{Experiments with different numbers of input prompt tokens to the local and global transformers (Please see the detailed definition in Sec~\ref{sec:loc-glo}) to justify our final choice. Using more prompt tokens means learning more parameters. We noticed that one prompt token for both local and global transformers is sufficient for achieving satisfactory performance.}\label{tab:prompt_token_num}
\centering
\resizebox{\columnwidth}{!}{
\begin{tabular}{cc|c|c|c}
\hline
\textbf{Num of input prompt tokens} &  \textbf{AUROC} & \textbf{Accuracy} & \textbf{Sensitivity} & \textbf{Specificity}\\
\hline
local: 2, global: 2 & 78.4 $\pm$ 4.2 &  69.5 $\pm$ 4.3 & \textbf{82.3 $\pm$ 9.9} & 55.2 $\pm$ 14.0\\
local:1, global: 1 & \textbf{79.6 $\pm$ 3.8} &  \textbf{72.5 $\pm$ 4.4} & 80.1 $\pm$ 5.8 & \textbf{65.7 $\pm$ 6.4}\\
\hline
\end{tabular}
}
\end{table}

\subsection{Attention Scores of ASD vs. HC in Comparison between Com-BrainTF (ours) and BNT (baseline)}
\begin{figure}[]
\centering
\includegraphics[width=12cm]{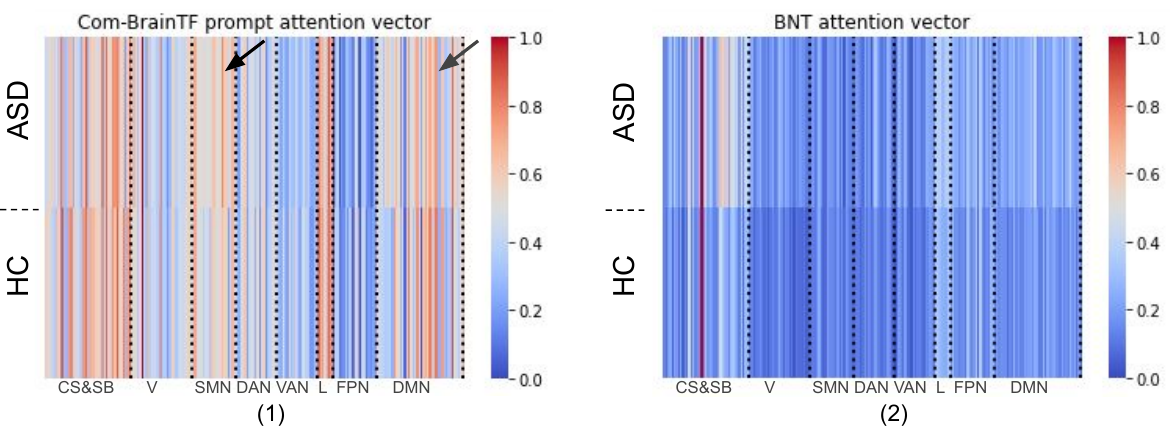}
\caption{Comparison of Com-BrainTF's prompt vector with BNT's attention vector to demonstrate improved interpretability of Com-BrainTF (1) Visualization of prompt's attention vector that shows most differences in DMN and SMN communities (2) Visualization of the sum of rows of the attention matrix in the baseline, BNT. In comparison to Com-BrainTF where clear  differences between ASD and HC are seen in DMN and SMN, we cannot tell the difference between ASD and HC groups using BNT.} \label{fig:prompt_attn_vis}
\end{figure}

\subsection{Decoded Functional Group Differences of ASD vs. HC}

\begin{figure}[]
\centering
\includegraphics[width=11cm]{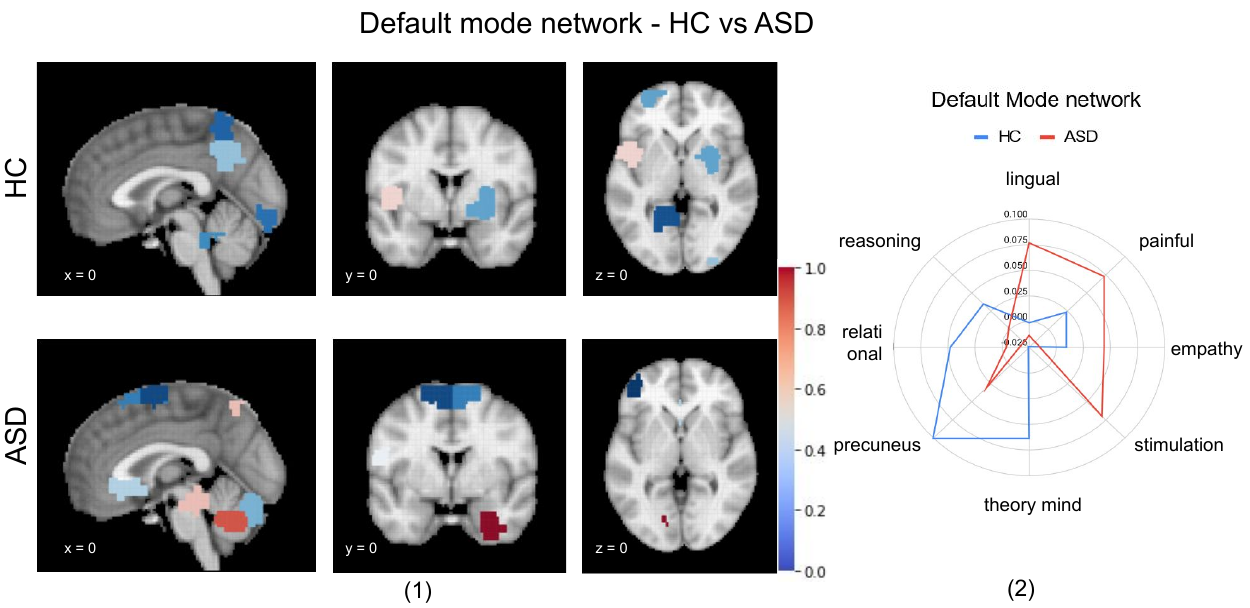}
\caption{We follow the analysis similar to Sec~\ref{sec:quant} but focus on the positive components of the normalized attention scores differences between ASD vs HC. Namely, we analyze the average $score_{\rm ASD} = (attn_{\rm ASD} - attn_{\rm HC})_{+}$ for ASD and $score_{\rm HC} = (attn_{\rm HC} - attn_{\rm ASD})_{+}$ for HC. (1) Visualization of $score_{\rm ASD}$ and $score_{\rm HC}$ showing differences in HC and ASD subjects in the DMN functional community. (2) Correlation between highlighted ROIs of ASD and HC in DMN community, and functional keywords decoded by Neurosynth. The interpreted results indicate functional differences between ASD and HC, potentially providing new insights for discovering ASD biomarkers.} \label{fig:DMN}

\end{figure}

\begin{figure}[]
\centering
\includegraphics[width=11cm]{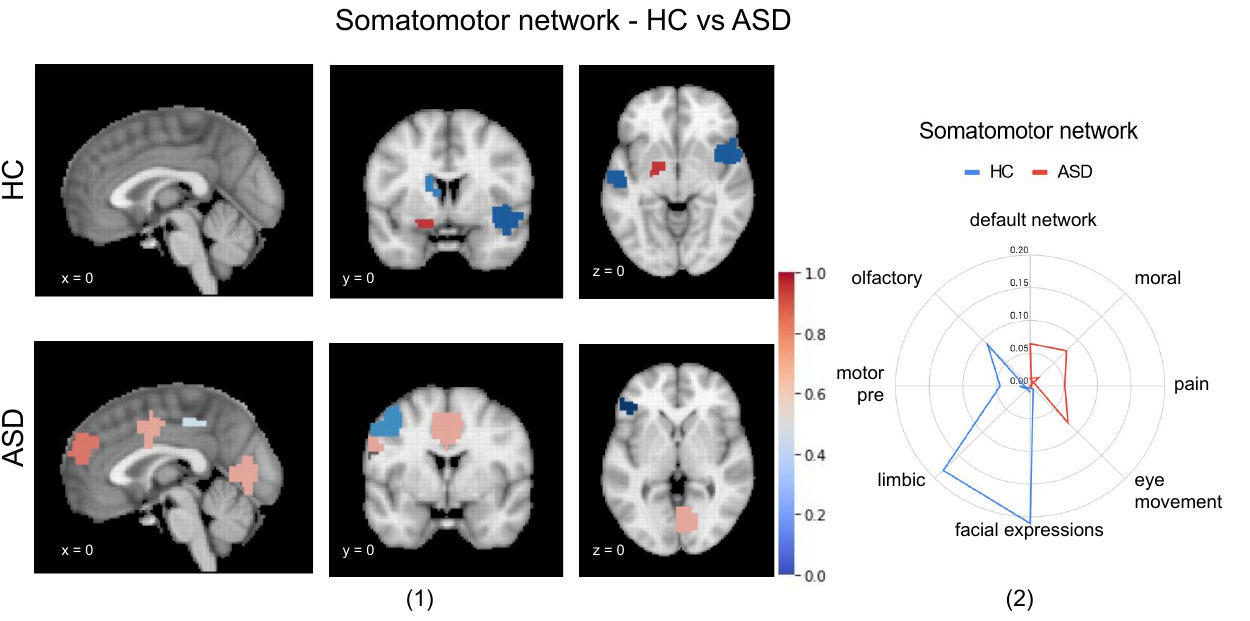}
\caption{We follow the analysis similar to Sec~\ref{sec:quant} but focus on the positive components of the normalized attention scores differences between ASD vs HC. Namely, we analyze the average $score_{\rm ASD} = (attn_{\rm ASD} - attn_{\rm HC})_{+}$ for ASD subjects and $score_{\rm HC} = (attn_{\rm HC} - attn_{\rm ASD})_{+}$ for HC subjects. (1) Visualization of $score_{\rm ASD}$ and $score_{\rm HC}$ showing differences in HC and ASD subjects in the SMN functional community. (2) Correlation between highlighted ROIs of ASD and HC in SMN community, and functional keywords decoded by Neurosynth. The interpreted results indicate functional differences between ASD and HC, potentially providing new insights for discovering ASD biomarkers.} \label{fig:SMN}
\end{figure}

\end{document}